\documentclass[aip,amsmath,amssymb,reprint]{revtex4-2}
\usepackage{graphicx}% Include figure files
\usepackage[utf8]{inputenc}
\usepackage[T1]{fontenc}
\usepackage[pdfstartview=FitH, pdfpagemode=None, pdfpagelayout=OneColumn, colorlinks=true, linkcolor=blue, citecolor = blue]{hyperref}

\begin{document}

\title{Acceleration rate enhancement by negative plasma density gradient in multi-bunch driven plasma wakefield accelerator}

\author{N.V. Okhotnikov and K.V. Lotov}
\affiliation{Novosibirsk State University, Novosibirsk 630090, Russia}
\affiliation{Budker Institute of Nuclear Physics, Novosibirsk 630090, Russia}
\date{\today}
\begin{abstract}
In a plasma wakefield accelerator driven by a train of short particle bunches, it is possible to locally increase the acceleration rate by introducing a small negative gradient of the plasma density. 
A regime is possible in which the gradient affects only the relative phasing of the driver bunches and the wave, keeping the wave phase behind the driver stable. 
With this technique, it is possible to increase the energy gain of the accelerated witness bunch in a plasma section of limited length.
\end{abstract}

\maketitle
The acceleration of particles in plasmas, in particular plasma wakefield acceleration, offers a path to higher particle energies and smaller facility sizes compared to conventional accelerators, and is being actively studied in this context.\cite{PoP27-070602,NJP23-031101}
The wave (called wakefield) that accelerates particles is most often excited (or driven) by a single short laser pulse\cite{APB74-355} or particle bunch.\cite{PRA44-6189}
Resonant wakefield excitation by a sequence of short laser pulses or particle bunches, which was envisioned in the original proposals for plasma-based accelerators,\cite{JETPL13-354,PRL43-267,PRL54-693} is less common but still of interest because it is sometimes easier to create a train of weak drivers\cite{JPB47-234003,PRL104-255003} than a single strong driver.

For proton-driven plasma wakefield acceleration,\cite{RAST9-85} which may prove to be the shortest path to the highest energies of accelerated leptons,\cite{Symmetry14-1680,PoP18-103101,PPCF63-125027,JINST17-T05008} the multi-bunch drive is a must.
This is because proton beams in the available synchrotrons are much longer than the plasma wavelength and cannot excite the wave directly.
Their frequency spectrum contains almost no harmonics at the plasma frequency for plasma densities required for strong accelerating gradients, so the beams must be either compressed in the longitudinal direction or modulated.
Compressing a high-energy proton beam to a single sub-millimeter-long bunch, which is necessary for efficient beam-plasma interaction, is prohibitively difficult.\cite{PAC09-4542,IPAC10-4395,PPCF53-014003} 
Therefore, it is more practical to convert a proton beam into a bunch train either in the plasma\cite{PRL104-255003} or before entering the plasma.\cite{RuPAC16-303}

Wave excitation by trains of particle bunches has been and is being studied in the AWAKE experiment\cite{NIMA-829-76, Nat.561-363, PPCF60-014046, Symmetry14-1680} at CERN and in several experiments with electron beams around the world.\cite{PPR20-596, NIMA-292-12, NIMA-829-334, PRAB24-051302, PRL112-045001, PRL120-144802, PPCF61-045012, PoP29-100701}
The measurements are in good agreement with numerical simulations,\cite{PPCF62-125023, PRAB23-081302, PRAB24-011301, PRAB24-101301} despite the fact that multi-bunch drivers are far away from other plasma wakefield accelerators in parameter space.\cite{NIMA-909-446}
Thus, the physics of beam-plasma interaction in the multi-bunch regime is well described by numerical codes, and their predictions are reliable.

To efficiently excite the wakefield over long distances, the bunches of the drive beam must be completely in the decelerating and focusing phases of the wave.
For this, the bunch-to-bunch spacing must be slightly longer than the plasma wavelength and dependent on the bunch position in the train.\cite{AIP396-75, PPCF60-024002}
Such a bunch train can be produced not only by manually positioning individual bunches,\cite{PRL101-054801, PRST-AB13-052803, NIMA-865-139} but also in plasma as a result of seeded self-modulation.\cite{PoP18-024501, PoP22-103110, PoP20-103111, JPCS1067-042009}
The bunch positioning in the latter case is controlled by the longitudinal profile of the plasma density at the self-modulation stage.

It is now generally understood what density profile is required to produce a bunch train that propagates stably in a uniform plasma and excites a strong wakefield until longitudinal effects such as beam depletion or dephasing come into play.
This profile must have a ``density step'', i.e., a sharp or smooth increase in the plasma density at the linear stage of beam self-modulation.\cite{PoP22-103110, PoP18-103101, PPCF63-125027}
The effect of weak density gradients on self-modulation was also studied both theoretically and experimentally, with the conclusion that small positive gradients enhance the witness energy gain relative to the uniform plasma.\cite{PRL107-145003, NIMA-829-3, NIMA-829-63, PPCF61-104004, Nat.561-363, PTRSA377-20180418, PPCF62-125023, PRL125-264801, PRAB24-101301}
Uniform density plasmas are far from the best self-modulation scenario, because the beam transforms into an unstable bunch train that is destroyed by its own wakefield in a time comparable to the self-modulation time.\cite{PoP18-024501, PoP18-103101, PoP20-103111, PoP22-103110, PoP22-123107,Symmetry14-1680}
Nevertheless, at short interaction distances, uniform plasmas can demonstrate high acceleration rates and serve as a proof of the self-modulation concept.\cite{Nat.561-363, PRL122-054802, PRL122-054801}

While perspective applications of multi-bunch wakefield excitation require as stable bunches as possible,\cite{PPCF63-125027, PTRSA377-20180185, Symmetry14-1680} near-future experiments\cite{Symmetry14-1680} may have a different target function and aim to obtain the highest witness energy within a limited acceleration distance.
In this paper, we show that if the acceleration distance is limited and much shorter than the beam dephasing length, the witness energy gain can be increased with a negative gradient of the plasma density.
The non-trivial point here is that it is possible to select such parameters of the system that the density gradient, affecting the driver and the wave amplitude, does not change the phase velocity of the wave at the witness location, and particles can be accelerated in such a wave.

\begin{table}[tb]
\begin{center}
\caption{Parameters of the system}\label{t1}
\begin{tabular}{llc}\hline\hline
  Parameter, notation && Value \\ \hline
  \textit{Plasma:} && \\
  Initial density, $n_0$ && $2 \times 10^{14}\,\text{cm}^{-3}$ \\
  Density after rise, $n_1$ && $2.11 \times 10^{14}\,\text{cm}^{-3}$ \\
  Length of plasma sections, $L_1$, $L_2$  && 10\,m \\
  Length of vacuum gap, $L_v$ && 1\,m \\
  Length of density ramp, $L_s$ && 5.2\,m \\
  Length of vacuum-plasma transition,\cite{PPCF62-115025} $D$ && 1\,cm \\
  Radius && 1.4\,mm \\
  Ion mass number && 85 \\
  \textit{Proton beam:} && \\
  Population, $N_b$ && $3\times 10^{11}$ \\
  Length, $\sigma_{zb}$ && 7.5\,cm \\
  Radius, $\sigma_{rb}$ && 0.2\,mm \\
  Energy && 400\,GeV \\
  Energy spread && 0.035\,\% \\
  Normalized emittance && 3.6\,mm\,mrad \\
  \textit{Seed electron beam:} && \\
  Population, $N_e$ && $3.125\times 10^9$ \\
  Length, $\sigma_{ze}$ && 0.714\,mm \\
  Radius, $\sigma_{re}$ && 0.19\,mm \\
  Energy && 18\,MeV \\
  Energy spread && 0\,\% \\
  Normalized emittance && 2.96\,mm\,mrad \\
  Distance to proton beam centroid && 18.6\,cm \\
  \textit{Simulations:} && \\
  Grid steps, $\Delta r$, $\Delta \xi$ && $0.01 k_p^{-1}$ \\
  Step for updating plasma state, $\Delta z$ && $200 k_p^{-1}$ \\
  Radius of simulation domain && $30 k_p^{-1}$ \\ \hline
  \hline
 \end{tabular}\end{center}
\end{table}

\begin{figure}[tb]
\includegraphics{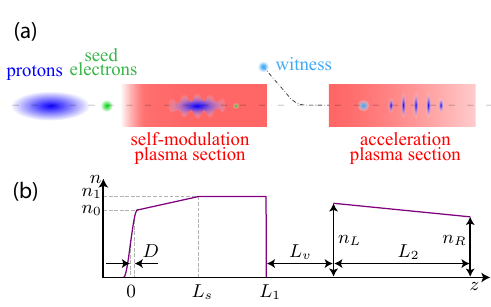}
\caption{(a) Plasma configuration under study and (b) longitudinal profile of the plasma density (not in scale).}\label{fig1-setup}
\end{figure}

Let us illustrate the effect with parameters achievable in the Run2c phase\cite{Symmetry14-1680} of the AWAKE experiment (Table~\ref{t1}).
The proton beam consecutively passes through two plasma sections [Fig.\,\ref{fig1-setup}(a)].
In the first section, the beam self-modulates. 
The formed bunches are phase-locked to the seed perturbation created by a low-energy electron bunch.
In the vacuum gap between plasma sections, remnants of the seed electron bunch are deflected from and a high-quality witness bunch is brought to the axis by a transverse magnetic field.
The witness is then accelerated in the second plasma section by the wakefield of the modulated proton beam.

The plasma density profile in the first section contains the rise of length $L_s$ necessary to form a stable bunch train [Fig.\,\ref{fig1-setup}(b)]. 
Its parameters were optimized similarly to Ref.~\onlinecite{PPCF62-115025} but for two-sectioned plasma.
The choice of a relatively low plasma density $n_0$ (Table~\ref{t1}) is convenient in view of possible experimental verification, since it facilitates beam alignment and diagnostics.\cite{PRL129-024802, PRL132-075001}
The second plasma section in AWAKE allows constant density gradients,\cite{JPD51-025203} so we vary the densities $n_L$ and $n_R$ at its ends [Fig.\,\ref{fig1-setup}(b)] in search of maximum witness energy.

We will not focus on the exact matching of wave and witness, as this requires additional multi-parameter optimization.
Instead, we approximate the witness energy gain using the distributions of accelerating and focusing wakefields on the axis.
This method can typically predict the maximum energy gain of a real witness with an accuracy of about 10\%.\cite{PPCF63-125027, BLPI50-S715, PPNL21-316}

We simulate the beam dynamics with axisymmetric quasistatic code LCODE\cite{PRST-AB6-061301,NIMA-829-350} that calculates all quantities as functions of radius $r$, longitudinal coordinate $z$ and co-moving coordinate $\xi=z-ct$, where $t$ is time and $c$ is the speed of light. 
The code uses dimensionless units, so we measure densities in units of $n_0$, times in units of $\omega_p^{-1}$ and distances in units of $k_p^{-1} = c/\omega_p$, where $\omega_p = \sqrt{4 \pi n_0 e^2/m}$ is the electron plasma frequency, $e$ is the elementary charge, and $m$ is the electron mass.
Typical run parameters are given in Table~\ref{t1}.
At this resolution, simulation of a single variant takes about 300 CPU hours.

\begin{figure}[tb]
\includegraphics{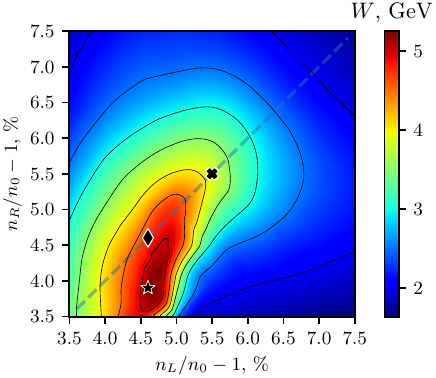}
\caption{Maximum witness energy gain $W$ as a function of plasma densities $n_L$ and $n_R$ at the ends of the 2nd plasma section. 
The cross, diamond, and asterisk mark the baseline, flat, and optimum variants, respectively.}
\label{fig2-grid}
\end{figure}

\begin{figure}[tb]
\includegraphics{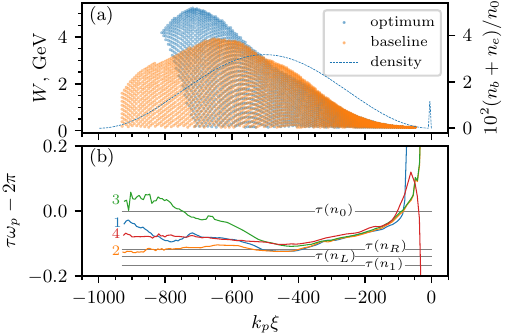}
\caption{(a) Dependence of the witness energy gain $W$ on witness position $\xi$ for the baseline and optimum variants. 
The dashed line shows the density of proton ($n_b$) and seed electron ($n_e$) beams on the axis before entering the 1st plasma section.
(b) Wakefield period $\tau$: at the end of the 1st plasma section (1), at the beginning of the 2nd plasma section for the baseline (2) and optimum (3) variants, and at the end of the 2nd plasma section for the optimum variant (4). 
The thin horizontal lines show the period of free plasma oscillations for the plasma densities written over the lines.}\label{fig3-distribution}
\end{figure}

The witness gains the highest energy if the plasma density in the second section is lower than the density at the end of the first section ($n_L < n_1$), and there is a negative density gradient ($n_L > n_R$) (Fig.\,\ref{fig2-grid}). 
For our particular parameters, the highest energy is achieved for $n_L = 1.046 n_0$, $n_R = 1.039 n_0$.
We call this variant ``optimum''  in contrast to the ``baseline'' variant with $n_L = n_R = n_1$ and the ``flat'' variant with $n_L = n_R = 1.046 n_0$.
The energy gain in the optimum variant is 33\% higher that in the baseline variant and is reached for a later witness location [Fig.\,\ref{fig3-distribution}(a)].

To clarify the reasons for this energy behavior, let us analyze the changes in the wakefield phase and bunch shapes as the proton beam propagates through the second plasma section. 
We characterize the wakefield phase by locations $\xi_\text{loc}$ of zero-field points on the axis, where $E_z = 0$ and $\partial E_z / \partial \xi > 0$.
For visualizing the longitudinal profile of the bunches, we use the dimensionless effective current\cite{PPCF64-075003}
\begin{equation}\label{er1}
    I_\text{eff} (\xi, z) = 4 \pi r_e \int_0^\infty K_0 (k_p r) n_b (r, \xi, z) r \, dr,
\end{equation}
where $r_e$ is the classical electron radius, $K_0$ is the modified Bessel function, and $n_b$ is the beam density. 

\begin{figure*}[tb]
\includegraphics{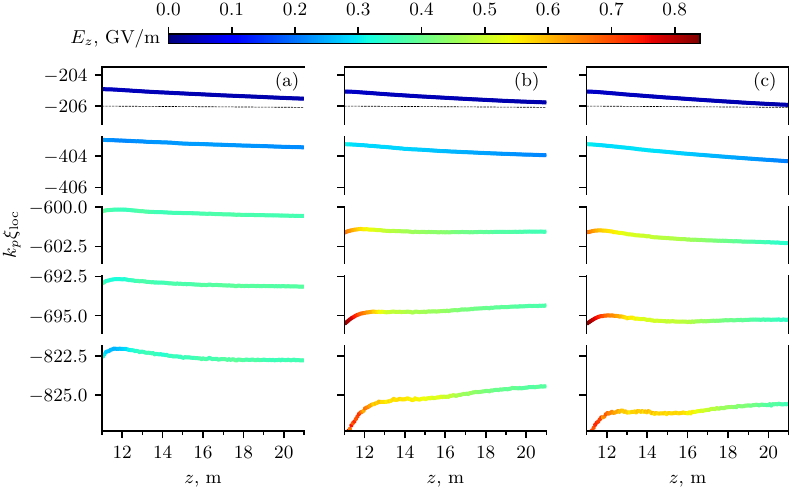}
\caption{Lines of constant wave phase for the baseline (a), flat (b), and optimum (c) variants in the 2nd plasma section. 
Only selected lines from different parts of the beam are shown. 
The line color shows the local wakefield amplitude. 
The almost invisible slope of thin dashed lines corresponds to the velocity of 400\,GeV protons.}\label{fig4-phases}
\end{figure*}

In the region of well-formed proton bunches at the leading half of the beam (at $100 \lesssim |k_p \xi| \lesssim 500$), the wave is phase-locked to the bunches, and its period is close to the bunch-to-bunch distance for all examined variants [Fig.\,\ref{fig3-distribution}(b)].
The period is longer than the period of free plasma oscillations for the corresponding plasma densities and slightly decreases along the beam, which is typical for optimally formed bunch trains.\cite{AIP396-75, PPCF60-024002}
This can be imagined as each bunch lengthens the period of the wave with its contribution.\cite{PoP22-103110, PPCF60-024002}
The phase velocity of the wave here is less than the velocity of 400\,GeV protons (Fig.\,\ref{fig4-phases}) because the bunches change their shape.
Namely, the self-modulation does not complete within the first plasma section and continues in the second section, reducing the phase velocity there.\cite{PRL107-145003, PRL107-145002}
When the bunches are fully formed, the phase velocity approaches the velocity of beam particles, but this becomes visible at longer propagation distances.\cite{PPCF63-125027}

\begin{figure}[tb]
\includegraphics{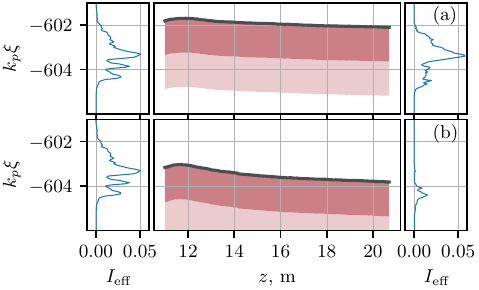}
\caption{Lines of constant wave phase that correspond to local maxima of decelerating field (black, in the middle column) and effective beam current $I_\text{eff}$ in their vicinity at the entrance to (left column) and exit from (right column) the 2nd plasma section for the baseline (a) and optimum (b) variants. 
Shaded areas show the quarters of the wave period in which the proton beam is focused and either decelerated (darker) or accelerated (lighter).}
\label{fig5-driv-evol}
\end{figure}
\begin{figure}[tb]
\includegraphics{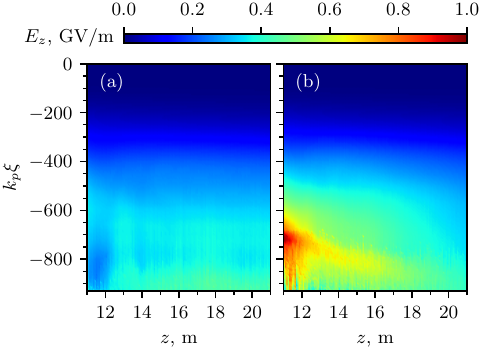}
\caption{Wakefield amplitude $E_z(z,\xi)$ for baseline (a) and optimum (b) variants.}\label{fig6-elocf}
\end{figure}

In the baseline variant, the bunches perfectly fit the focusing wave phase, and hence the bunch train is stable.
The beam remains stable even after passing the gap between the plasma sections, despite the radial expansion of the bunches, the corresponding decrease in the bunch density and excited wakefield,\cite{Symmetry14-1680} and the shortening of the wakefield period at the rear part of the beam [compare lines 1 and 2 in Fig.\,\ref{fig3-distribution}(b)].
As for wave driving, not all bunches are equally efficient. 
The self-modulation develops in such a way\cite{PoP22-103110} that the last bunches of the train are partly in the decelerating phase [Fig.\,\ref{fig5-driv-evol}(a)] and therefore less efficient.
If the plasma density at the beginning of the second section is lower than in the baseline variant, the wakefield period elongates [compare lines 2 and 3 in Fig.\,\ref{fig3-distribution}(b)], the wave shifts backward relative to the bunches [compare Fig.\,\ref{fig4-phases}(a) and (b) at $|k_p \xi_\text{loc}| > 500$], and tail bunches fall into the phase of stronger decelerating field [Fig.\,\ref{fig5-driv-evol}(b)].
With this phasing, the bunches drive the wave more efficiently, and the wave amplitude increases compared to the baseline variant (Fig.\,\ref{fig6-elocf} and line colors in Fig.\,\ref{fig4-phases}).

\begin{figure}[tb]
\includegraphics{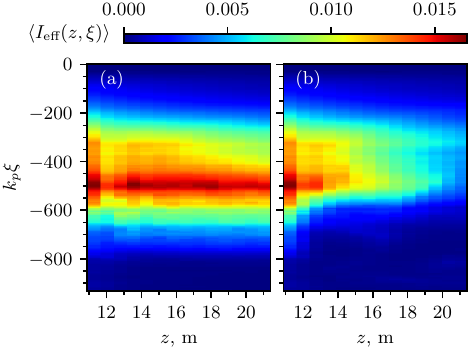}
\caption{Effective current averaged over the wakefield period $\langle I_\text{eff}(z,\xi) \rangle$ for baseline (a) and optimum (b) variants.}\label{fig7-ieff}
\end{figure}

In the plasma of reduced density, some parts of the bunches fall into the defocusing wave phase or into a weaker focusing phase. 
In both cases, they gradually deteriorate [Fig.\,\ref{fig5-driv-evol}(b)]. 
The defocusing force is strongest in the beam tail, so the tail bunches disappear earlier (Fig.\,\ref{fig7-ieff}).
The destroyed bunches can no longer lengthen the wakefield period, so the period shortens [compare lines 3 and 4 in Fig.\,\ref{fig3-distribution}(b) at $|k_p \xi| \gtrsim 500$]. 
This makes the wave behind the bunches superluminal and unsuitable for witness acceleration [region $|k_p \xi_\text{loc}| \gtrsim 700$ in Fig.\,\ref{fig4-phases}(b)].
Note that the wakefield period behind the driver does not tend to the period of free plasma oscillations [Fig.\,\ref{fig3-distribution}(b)]. 
There are two reasons for this: the disturbance of the ion background,\cite{PRL109-145005, PoP21-056705} leading to a local decrease of the plasma frequency, and the Doppler effect due to a negative average velocity of plasma electrons, which they acquire when interacting with the beam.\cite{PRL112-194801, PPCF64-045003}

As we see, there is a transition between subluminal wave (at small $|\xi|$) and superluminal wave (at larger $|\xi|$), at which the wave is best suited for witness acceleration. 
In the ``flat'' variant, this occurs at $|k_p \xi| \approx 600$ [Fig.\,\ref{fig4-phases}(b)]. 
Plasma density gradients can control the location of this transition point.
The negative gradients act on the wave similarly to the density decrease at the beginning of the second plasma section, but in a time-distributed manner.
It shifts the wave backward relative to the bunches. 
In the ``optimum'' variant, the line of constant phase flattens in the region of the strongest accelerating field [at $|k_p \xi| \approx 700$ in Fig.\,\ref{fig4-phases}(c)], providing the highest energy gain [Fig.\,\ref{fig3-distribution}(a)].

To summarize, by varying the longitudinal profile of the plasma density, the wakefield phase can be controlled and optimized for either the longest or highest-rate acceleration.
In the case of long bunch trains, the relationship between plasma density and wave phase is not as explicit as for short drivers, but with numerical simulations it can be elucidated and utilized.
With a negative density gradient introduced after the beam self-modulation, it is possible to significantly increase the witness energy gain in a fixed-length plasma section. 
The non-trivial point here is that the wakefield phase shift, which initiates the rapid driver destruction and the increase in wakefield amplitude, does not affect the accelerated beam.
This effect is unlikely to be useful for energy enhancement in optimally designed plasma wakefield accelerators, where the interaction length is determined by bunch depletion or dephasing. 
However, in experiments with short plasmas, it can be reliably diagnosed by the witness energy gain, which will show either a good understanding of the underlying processes (in case of agreement with simulations) or the presence of missing physical effects.

\acknowledgements

This study was supported by the Russian Science Foundation, Project No. 23-12-00028. 
Optimization of the plasma density profile in the first plasma section [Fig.\,\ref{fig1-setup}(b)] was performed on HPC-cluster ``Akademik V M Matrosov''\cite{Matrosov} and supported by Budker INP with ongoing institutional funding. 
The authors thank Vlada Yarygova for help in optimizing the plasma density profiles.

\end{document}